\documentclass[conference]{IEEEtran}
\IEEEoverridecommandlockouts
\usepackage{cite}
\usepackage{amsmath,amssymb,amsfonts}
\usepackage{algorithmic}
\usepackage{graphicx}
\usepackage{textcomp}
\usepackage{xcolor}
\usepackage{booktabs}
\usepackage{multirow}
\def\BibTeX{{\rm B\kern-.05em{\sc i\kern-.025em b}\kern-.08em
    T\kern-.1667em\lower.7ex\hbox{E}\kern-.125emX}}

\usepackage{marvosym}  

\begin{document}

\title{Reliability-Aware Geometric Fusion for Robust Audio-Visual Navigation}

\author{
Teng Liu$^{1,2,3}$, and Yinfeng Yu$^{1,2,3}$$^{,\mbox{\Letter}}$%
\thanks{$^{\mbox{\Letter}}$Yinfeng Yu is the corresponding author(E-mail: yuyinfeng@xju.edu.cn).}%
\\
$^1$Joint Research Laboratory for Embodied Intelligence, Xinjiang University\\
$^2$Joint International Research Laboratory of Silk Road Multilingual Cognitive Computing, Xinjiang University\\
$^3$School of Computer Science and Technology, Xinjiang University, Urumqi 830017, China%
}

\maketitle

\begin{abstract}
Audio-Visual Navigation (AVN) requires an embodied agent to navigate toward a sound source by utilizing both vision and binaural audio. A core challenge arises in complex acoustic environments, where binaural cues become intermittently unreliable, particularly when generalizing to previously unheard sound categories. To address this, we propose RAVN (Reliability-Aware Audio-Visual Navigation), a framework that conditions cross-modal fusion on audio-derived reliability cues, dynamically calibrating the integration of audio and visual inputs. RAVN introduces an Acoustic Geometry Reasoner (AGR) that is trained with geometric proxy supervision. Using a heteroscedastic Gaussian NLL objective, AGR learns observation-dependent dispersion as a practical reliability cue, eliminating the need for geometric labels during inference. Additionally, we introduce Reliability-Aware Geometric Modulation (RAGM), which converts the learned cue into a soft gate to modulate visual features, thereby mitigating cross-modal conflicts. We evaluate RAVN on SoundSpaces using both Replica and Matterport3D environments, and the results show consistent improvements in navigation performance, with notable robustness in the challenging unheard sound setting.
\end{abstract}


\section{Introduction}

Navigation in the physical world is inherently multi-modal. Humans use vision to interpret detailed scenes, while relying on audition to detect events outside their direct line of sight. Audio-Visual Navigation (AVN) seeks to replicate this ability in embodied agents, guiding them toward sound sources in complex indoor environments~\cite{chen2020soundspaces}. However, in realistic settings, sound waves are often distorted by obstacles, diffraction, and scene geometry, making binaural localization cues intermittently unreliable. Consequently, the potential for precise metric localization in such chaotic acoustic environments is inherently limited. This challenge intensifies when agents must generalize to novel sound sources. Unheard sounds induce a distribution shift in audio features, complicating the extraction of reliable binaural cues and increasing the risk of following spurious patterns or drifting in incorrect directions~\cite{yu2023echo,yu2025dgfnet,zhang2025advancing}.
\begin{figure}[t] 
  \centering
\includegraphics[width=0.88\linewidth]{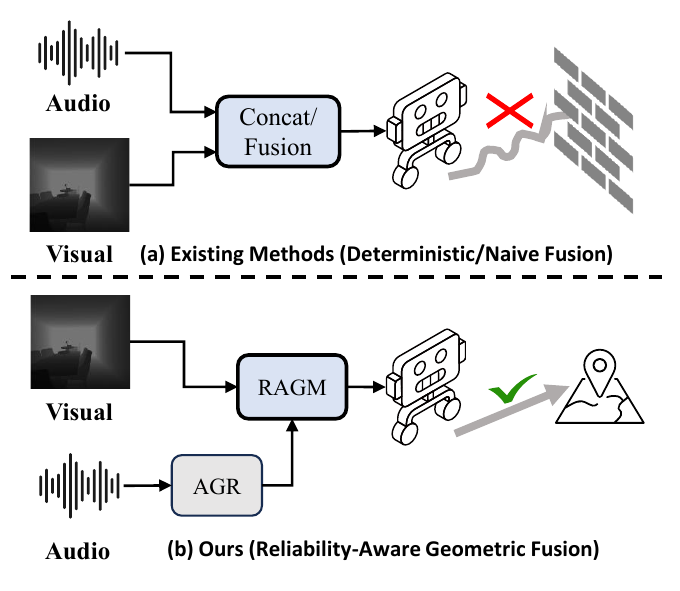}
 \caption{\textbf{Paradigm comparison.} (a) naive deterministic fusion vs. (b) our reliability-aware RAVN framework.}
  
  \label{fig:gn} 
\end{figure}

\begin{figure*}[t] 
  \centering
  \includegraphics[width=0.90\linewidth]{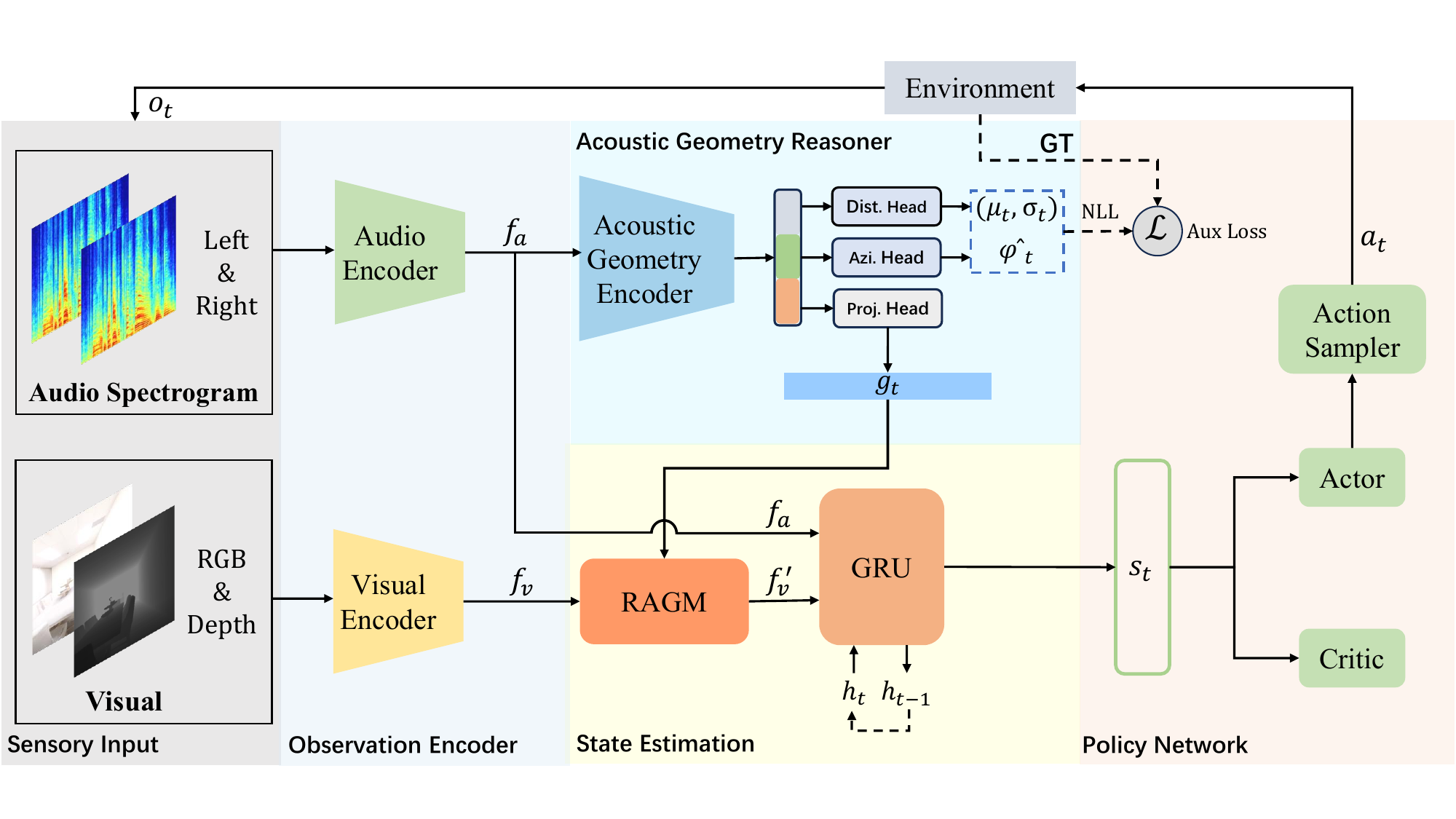}
\caption{\textbf{The RAVN Framework.} Multi-modal observations $o_t$ are encoded into $f_a$ and $f_v$. The AGR module estimates geometric embeddings $g_t$ and predictive uncertainty ($\mu_t, \sigma_t$), supervised by Ground Truth (GT) labels via an auxiliary loss. The RAGM module then uses $g_t$ to dynamically modulate visual features into $f'_v$. A recurrent policy (GRU) aggregates these components into the hidden state $s_t$ for end-to-end action ($a_t$) selection.}

  \label{fig:main} 
\end{figure*}

This motivates the need for a critical yet underexplored capability in AVN: reliability-aware geometric fusion—dynamically modulating cross-modal integration based on inferred geometric reliability. This approach fully leverages audio information when cues are consistent. It reduces the influence of audio when the evidence becomes ambiguous. Consider human intuition: when hearing a distant or reverberant sound, we don't blindly follow it. Instead, we instinctively assess the sound's reliability before deciding how much to trust it. If the sound is unclear, we reduce the impact of auditory cues and prioritize visual confirmation~\cite{garg2023visually}.

Most existing AVN methods lack such mechanisms. First, they implicitly assume a high degree of signal reliability, prioritizing precise geometric regression while underestimating the stochastic nature of distorted audio, and rely on naive concatenation for modality fusion, making policies vulnerable to acoustic illusions in distorted environments~\cite{zhang2025iterative,yu2025dynamic,li2025audio}. Second, while auxiliary tasks are sometimes used for representation learning, their predictions are generally limited to training objectives and don't directly inform fusion at test time. As a result, agents struggle to adaptively calibrate their trust in audio under varying environmental conditions.

To address this gap and inspired by human intuition, we propose Reliability-Aware Audio-Visual Navigation (RAVN), as conceptually contrasted with existing methods in Fig.~\ref{fig:gn}. Unlike static reliability assumptions, RAVN learns reliability-aware cues from binaural observations to modulate visual features, effectively filtering out unreliable auditory guidance and mitigating cross-modal conflicts. By down-weighting misleading cues and leveraging consistent evidence, it dynamically adjusts the fusion process, enabling robust navigation in complex acoustic environments.

In summary, we present the following contributions:
\begin{itemize}
\item We introduce RAVN, a reliability-aware framework that combines proxy geometric supervision with a heteroscedastic objective to learn audio-derived reliability cues.
\item We propose RAGM, a mechanism that uses these learned cues to modulate visual features, adjusting them based on audio reliability and mitigating cross-modal conflicts.
\item We provide thorough empirical results showing that modeling sensory reliability improves navigation performance, especially in challenging scenarios with unheard sound sources.
\end{itemize}

\section{Related Work}

Embodied visual navigation, predominantly driven by reinforcement learning in simulators like Habitat~\cite{savva2019habitat}, is inherently constrained by limited fields of view. This motivates Audio-Visual Navigation (AVN), which incorporates omnidirectional binaural cues to guide agents toward sound-emitting targets~\cite{chen2020soundspaces,yu2025dope}. Existing AVN methods typically fuse modalities via static mechanisms like concatenation or attention~\cite{yu2022pay,yu2021weavenet,yusound}. However, these approaches implicitly assume constant auditory reliability. In complex acoustics, factors like reverberation and multipath propagation heavily distort binaural cues~\cite{brahmbhatt2018geometry}. Consequently, static fusion leaves agents vulnerable to over-trusting misleading audio, especially under distribution shifts from unheard sounds.
To address such perceptual ambiguity, uncertainty modeling—such as heteroscedastic regression—is widely used in general reinforcement learning~\cite{loquercio2020general,jaderberg2016reinforcement}. Yet, its direct integration into AVN fusion remains underexplored. While some AVN frameworks employ geometric auxiliary tasks for representation learning~\cite{chen2020learning,gea2025semantic}, they aim to enhance deterministic feature distinctiveness rather than quantify signal quality. Similarly, recent probabilistic AVN methods~\cite{gan2020look,yu2023measuring} heavily rely on precisely calibrated uncertainty, which often proves brittle under realistic acoustic shifts.
Our work bridges this gap by explicitly learning audio-derived reliability to dynamically condition cross-modal fusion. Instead of requiring fragile, fully calibrated probabilistic inference, RAVN repurposes geometric auxiliary tasks to capture heteroscedastic uncertainty (i.e., variance in distance and azimuth inference) as a latent discriminative signal. This allows the agent to adaptively down-weight ambiguous auditory cues and fully leverage consistent evidence, ensuring robust navigation in chaotic acoustic environments.

\section{The Proposed Approach}

\subsection{Overview}
We formalize the problem as a reinforcement learning task where the agent learns a policy to efficiently navigate toward an active acoustic source in an unknown environment. At each time step $t$, the agent receives a raw multi-modal observation $o_t$ comprising visual depth maps and binaural audio. To navigate robustly under complex acoustics, we propose RAVN (Reliability-Aware Audio-Visual Navigation), a framework that leverages explicitly estimated acoustic reliability to guide cross-modal fusion. As shown in Fig.~\ref{fig:main}, RAVN consists of three core components: (i) an Acoustic Geometry Reasoner (AGR) that extracts ambiguity-aware geometric embeddings and heteroscedastic uncertainty from audio features; (ii) a Reliability-Aware Geometric Modulation (RAGM) module that dynamically recalibrates visual features via soft-gating, conditioned on the learned reliability; and (iii) a recurrent policy network that aggregates the modulated features to update its hidden state representation $s_t$ for end-to-end action selection.

\subsection{Acoustic Geometry Reasoner (AGR)}
\label{sec:agr}
The Acoustic Geometry Reasoner (AGR) comprises a shared Acoustic Geometry Encoder and three parallel branches: a primary geometric projection head for feature alignment, and two auxiliary heads for representation learning. Formally, the encoder processes CNN-extracted audio features $f_a$ to extract spatial geometric cues (i.e., relative distance and azimuth) and intrinsic reliability information, generating a latent geometric embedding $z_{geo}$. Subsequently, the projection head maps $z_{geo}$ into a geometric representation $g_t$, aligning the acoustic embedding with the visual feature dimension to facilitate element-wise modulation.

Crucially, our goal is not precise localization, but representation learning that captures observation-dependent dispersion. Rather than seeking full calibration, the heteroscedastic objective yields practical reliability cues to implicitly condition the fusion for robust navigation.

\begin{figure}[t] 
  \centering
  \includegraphics[width=0.90\linewidth]{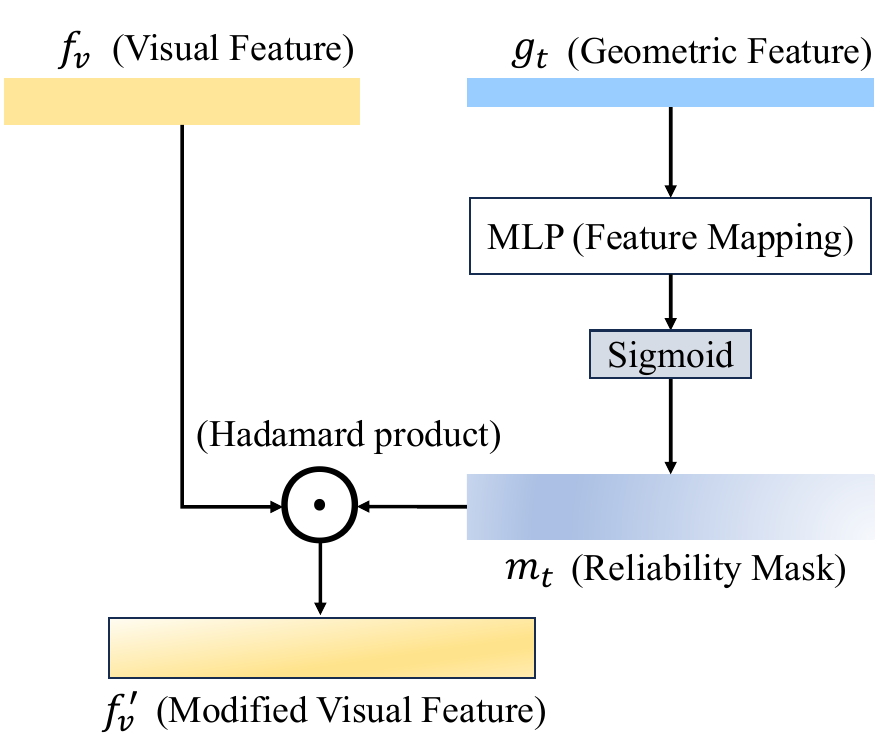}
 
\caption{\textbf{RAGM module architecture.} Geometric features are transformed into a reliability mask, which applies element-wise modulation to visual features to produce modified visual representations.}
  
  \label{fig:ragm} 
\end{figure}

\begin{table*}[t]
\centering
\caption{\textbf{Quantitative comparison on Replica and Matterport3D datasets.} We report Success Rate (SR), Success weighted by Path Length (SPL), and Success weighted by Number of Actions (SNA). Best results are highlighted in bold.}
\label{tab:main_results}

\resizebox{\linewidth}{!}{
\begin{tabular}{l|ccc|ccc|ccc|ccc}
\toprule

& \multicolumn{6}{c|}{\textbf{Replica}} & \multicolumn{6}{c}{\textbf{Matterport3D}} \\
\cmidrule(lr){2-7} \cmidrule(lr){8-13} 


\textbf{Method} & \multicolumn{3}{c|}{\textbf{Heard}} & \multicolumn{3}{c|}{\textbf{Unheard}} & \multicolumn{3}{c|}{\textbf{Heard}} & \multicolumn{3}{c}{\textbf{Unheard}} \\
\cmidrule(lr){2-4} \cmidrule(lr){5-7} \cmidrule(lr){8-10} \cmidrule(lr){11-13}

& SR $\uparrow$ & SPL $\uparrow$ & SNA $\uparrow$ & SR $\uparrow$ & SPL $\uparrow$ & SNA $\uparrow$ & SR $\uparrow$ & SPL $\uparrow$ & SNA $\uparrow$ & SR $\uparrow$ & SPL $\uparrow$ & SNA $\uparrow$ \\
\midrule

Random Agent~\cite{chen2020learning} & 18.5 & 4.9 & 1.8 & 18.5 & 4.9 & 1.8 & 9.1 & 2.1 & 0.8 & 9.1 & 2.1 & 0.8 \\
Frontier Waypoints~\cite{chen2020learning} & 63.9 & 44.0 & 35.2 & 14.8 & 6.5 & 5.1 & 42.8 & 30.6 & 22.2 & 16.4 & 10.9 & 8.1 \\
Direction Follower~\cite{chen2020learning} & 72.0 & 54.7 & 41.1 & 17.2 & 11.1 & 8.4 & 41.2 & 32.3 & 23.8 & 18.0 & 13.9 & 10.7 \\
Supervised Waypoints~\cite{chen2020learning} & 88.1 & 59.1 & 48.5 & 43.1 & 14.1 & 10.1 & 36.2 & 21.0 & 16.2 & 8.8 & 4.1 & 2.9 \\
Gan et al.~\cite{gan2020look} & 83.1 & 57.6 & 47.9 & 15.7 & 7.5 & 5.7 & 37.9 & 22.8 & 17.1 & 10.2 & 5.0 & 3.6 \\
AV-Nav ~\cite{chen2020soundspaces} & 93.0 & 76.1 & 44.7 & 47.3 & 36.8 & 21.5 & 68.8 & 52.3 & 29.6 & 34.5 & 25.0 & 12.7 \\

\midrule
\textbf{RAVN (Ours)} & \textbf{97.0} & \textbf{80.2} & \textbf{52.4} & \textbf{53.1} & \textbf{40.3} & \textbf{23.4} & \textbf{70.9} & \textbf{55.6} & \textbf{33.2} & \textbf{35.6} & \textbf{25.9} & \textbf{13.6} \\
\bottomrule
\end{tabular}
}
\end{table*}

\noindent\textbf{Probabilistic Distance Head:}
We model distance as a Gaussian $d_t \sim \mathcal{N}(\mu_t,\sigma_t^2)$ and optimize the Negative Log-Likelihood (NLL):
\begin{equation}
    \mathcal{L}_{dist} = \frac{1}{2}\log \sigma_t^2 + \frac{(y_{dist}-\mu_t)^2}{2\sigma_t^2},
\end{equation}
where $y_{dist}$ is the ground-truth geodesic distance.
The variance $\sigma_t^2$ captures aleatoric uncertainty, serving as a dynamic proxy for acoustic reliability.
Minimizing NLL forces $\sigma_t^2$ to increase when the prediction error is high, thereby explicitly signaling low reliability.

\noindent\textbf{Azimuth Head:}
Complementary to distance, we regress the relative azimuth $\hat{\varphi}_t$ to enforce orientation-aware feature learning.
To address angular periodicity (i.e., the discontinuity between $\pi$ and $-\pi$), we optimize the wrapped Smooth-L1 loss:
\begin{equation}
    \mathcal{L}_{ang} = \text{SmoothL1}\left(\mathcal{W}(\hat{\varphi}_t - y_{ang})\right),
\end{equation}
where $y_{ang}$ is the ground-truth azimuth and $\mathcal{W}(\cdot)$ wraps the error into $[-\pi,\pi]$.
This supervision ensures that the final embedding $g_t$ captures crucial directional semantics, providing the necessary spatial context to complement the distance-derived reliability.

\noindent\textbf{Projection Head:}
We project the latent embedding $z_{geo}$ into a geometric representation $g_t \in \mathbb{R}^{D_v}$ using a learnable mapping function $\phi_{proj}$:
\begin{equation}
g_t = \phi_{proj}(z_{geo}).
\end{equation}
This projection serves two main purposes: (1) it aligns the acoustic feature space with the visual features, enabling element-wise fusion, and (2) guided by auxiliary uncertainty gradients, $g_t$ encodes both spatial directionality and acoustic reliability, acting as a compact carrier for reliability-aware geometric reasoning.

\subsection{Reliability-Aware Geometric Modulation (RAGM)}
\label{sec:ragm}
The RAGM module converts learned reliability cues into an actionable fusion mechanism.
As detailed in Fig.~\ref{fig:ragm}, it takes the CNN-extracted visual feature $f_v$ and the geometric feature $g_t$ as inputs.
We generate a soft modulation mask $m_t \in (0,1)^{D_v}$ by applying an MLP followed by a sigmoid activation:
\begin{equation}
    m_t = \text{Sigmoid}(\text{MLP}(g_t)).
\end{equation}
The visual feature is then modulated via an element-wise Hadamard product:
\begin{equation}
    f'_v = f_v \odot m_t.
\end{equation}

Notably, the reliability mask $m_t$ serves as a step-by-step soft gate that is implicitly conditioned on acoustic ambiguity. Since $g_t$ shares its upstream encoder with the probabilistic heads, it captures latent noise-related cues through joint optimization. During navigation, $m_t$ dynamically fluctuates in response to the changing acoustic environment: it selectively re-weights visual channels by suppressing features that might lead to erroneous decisions when auditory evidence is unreliable, and enhancing them when the sound is clear. This element-wise modulation renders the fusion pathway temporally adaptive, allowing the agent to continuously recalibrate its multi-modal dependency under varying environmental complexity.

\subsection{Policy Learning via State Estimation}
\label{sec:policy}
We concatenate the raw audio and the modulated visual features as $x_t=[f_a, f'_v]$ and feed them into a GRU:
\begin{equation}
    s_t = \text{GRU}(x_t, h_{t-1}),
\end{equation}
where $h_t$ is the recurrent hidden state and $s_t$ is the state representation used by the actor and critic to produce $\pi(a_t|s_t)$ and $V(s_t)$.

Importantly, the modulated visual feature $f'_v$ is concatenated with the unmodulated audio feature $f_a$ before entering the GRU. While $f'_v$ adaptively gates visual exploratory cues based on environmental reliability, preserving the raw $f_a$ ensures the policy continuously receives the primary goal-directed auditory signal.

\subsection{Training Objectives}

The entire framework is trained end-to-end. While the primary policy is optimized via PPO~\cite{schulman2017proximal} to maximize navigation rewards, the robust representation learning in AGR is driven by Auxiliary Geometric Supervision ($\mathcal{L}_{aux}$).

Formally, we aggregate the supervision signals from the two auxiliary heads defined in Sec.~\ref{sec:agr}:
\begin{equation}
    \mathcal{L}_{aux} = \mathcal{L}_{dist} + \mathcal{L}_{ang},
\end{equation}
where $\mathcal{L}_{dist}$ is the heteroscedastic regression loss that instills uncertainty awareness, and $\mathcal{L}_{ang}$ denotes the cross-entropy loss for direction estimation.
The overall training objective is a weighted combination:
\begin{equation}
    \mathcal{L}_{total} = \mathcal{L}_{PPO} + \lambda \mathcal{L}_{aux},
\end{equation}
where $\lambda$ balances the reinforcement learning objective with geometric representation learning.

\begin{figure*}[t] 
  \centering
  \includegraphics[width=0.90\linewidth]{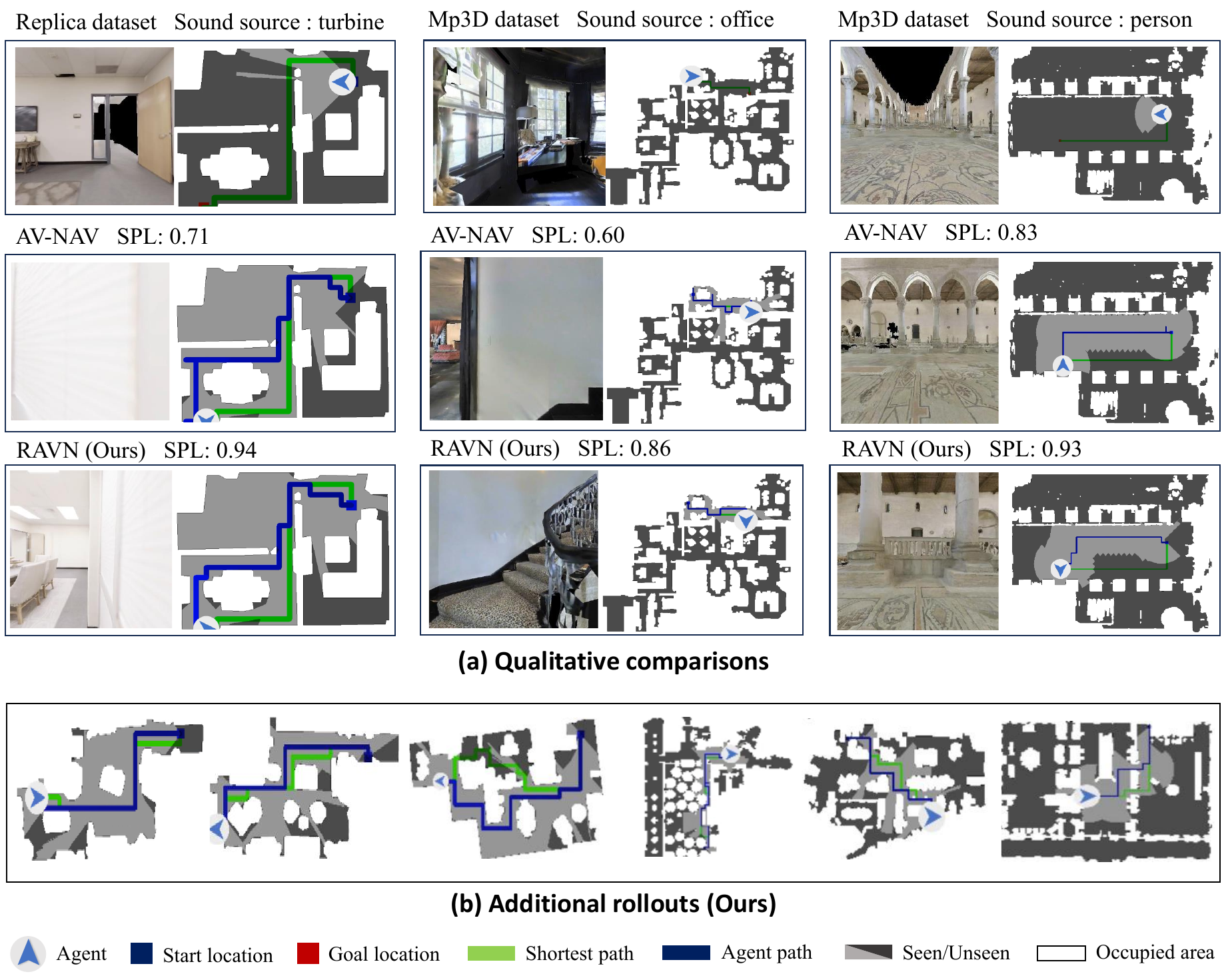}

\caption{\textbf{Qualitative results.} (a) Top-down trajectory comparisons between the AV-NaV baseline and our RAVN in representative Replica and Mp3D episodes, with SPL shown for each run. (b) Additional rollouts of RAVN across various layouts and start-goal configurations.}
  \label{fig:keshi} 
\end{figure*}

\section{Experiments}

\subsection{Experimental Setup} 
We implement our framework on the SoundSpaces platform using two publicly available datasets: Replica, which contains 18 scanned apartments and offices with 0.5m grids, and Matterport3D, comprising 85 large-scale home environments with 1m grids~\cite{chen2020soundspaces,straub2019replica,b22}. The RIR sampling rates are set to 44.1 kHz for Replica and 16 kHz for Matterport3D. We train the model using the Adam optimizer with a learning rate of $2.5 \times 10^{-4}$. Training runs for 48M steps on Replica and 60M steps on Matterport3D, with episodes limited to 500 steps. Following standard protocols, we evaluate two settings: (1) Multiple Heard: all 102 sounds are present in training, validation, and test splits; and (2) Multiple Unheard: sounds are divided into non-overlapping 73/11/18 splits to assess generalization.

\subsection{Evaluation Metrics} 
Following standard protocols, we employ three metrics to evaluate navigation performance: \textbf{Success Rate (SR)}, the fraction of episodes in which the agent reaches the goal within the time limit; \textbf{Success weighted by Path Length (SPL)}, our primary efficiency metric that weights success by the ratio of shortest-path (geodesic) distance to the executed path length, reflecting the ability to avoid acoustic detours; and \textbf{Success weighted by Number of Actions (SNA)}, which penalizes redundant actions and thus complements SPL in capturing decisiveness and stability under sensory ambiguity.

\subsection{Quantitative Experimental Results}
We adopt the task setup and evaluation protocol from SoundSpaces, with detailed results provided in Table~\ref{tab:main_results}.
On the Replica dataset, RAVN consistently outperforms the AV-Nav baseline across all sound conditions. In the challenging Unheard sound task, RAVN shows relative improvements of 12.3\% in SR and 9.5\% in SPL over the baseline. Furthermore, RAVN achieved a success rate of 53.1\% and a significantly higher path efficiency, highlighting improvements in cross-scenario generalization and navigation stability. In the Heard setting, RAVN maintains strong performance, achieving a near-perfect success rate of 97.0\%, while SNA improves by 17.2\%, demonstrating enhanced decisiveness in familiar acoustic environments.

On the more challenging Matterport3D dataset, RAVN continues to outperform the baseline across all sound conditions. In the Heard sound task, RAVN achieves a success rate of 70.9\%, with a 12.2\% relative improvement in SNA, indicating better navigation efficiency. In the difficult Unheard setting, RAVN achieves a 35.6\% success rate when navigating to unseen sound sources, demonstrating a 3.2\% relative improvement over the baseline, thereby further confirming the robust effectiveness of reliability-aware fusion in complex environments.

\subsection{Qualitative Experimental Results}

Fig.~\ref{fig:keshi}(a) shows top-down trajectories in Replica and Matterport3D environments, covering compact, cluttered, and open layouts. RAVN closely follows the geodesic shortest path, with fewer detours and less backtracking compared to the AV-Nav baseline—supporting our quantitative improvements in SPL.

The performance gap is most noticeable in regions prone to ambiguity, such as sharp corners, occluded doorways, and long corridors, where reverberation and multi-path effects distort binaural cues. While the baseline often becomes disoriented, drifting into unnecessary local exploration and requiring repeated re-orientations, RAVN progresses smoothly and decisively. This stability is attributed to our reliability-aware fusion design, which leverages audio-derived reliability cues learned via geometric proxy objectives. RAVN filters out misleading auditory cues that typically confuse standard agents.

Additional rollouts in Fig.~\ref{fig:keshi}(b) further demonstrate that this robustness holds across various scene layouts and start-goal configurations.

\subsection{Ablation Studies}
Tables~\ref{tab:ablation_replica} and~\ref{tab:ablation_mp3d} detail our ablation studies, specifically addressing module modification and deletion. Our baseline, AV-Nav [1], uses naive audio-visual concatenation without geometric or reliability modeling. Adding the Acoustic Geometry Reasoner with a deterministic loss (+AGR MSE) improves performance, validating geometry-based representation learning. Modifying this objective to a heteroscedastic Negative Log-Likelihood (+AGR NLL) yields further gains, demonstrating that capturing observation-dependent uncertainty is more effective than enforcing rigid metric accuracy under complex acoustics. Finally, the full RAVN model incorporates the RAGM module. Comparing RAVN to +AGR (NLL) directly isolates the impact of RAGM deletion; its removal causes significant performance drops, confirming that RAGM effectively translates learned reliability into an adaptive gate to filter misleading audio and ensure robust navigation.

\begin{table}[t]
\centering
\caption{\textbf{Ablation Study on Replica Dataset.} We compare different variants to validate the effectiveness of our proposed modules.}
\label{tab:ablation_replica}
\resizebox{\linewidth}{!}{

\begin{tabular}{l|ccc|ccc}
\toprule

& \multicolumn{6}{c}{\textbf{Replica}} \\
\cmidrule(lr){2-7}

\textbf{Model Variant} & \multicolumn{3}{c|}{\textbf{Heard}} & \multicolumn{3}{c}{\textbf{Unheard}} \\
\cmidrule(lr){2-4} \cmidrule(lr){5-7}


& SR($\uparrow$) & SPL($\uparrow$) & SNA($\uparrow$) & SR($\uparrow$) & SPL($\uparrow$) & SNA($\uparrow$) \\
\midrule

AV-Nav~\cite{chen2020soundspaces}  & 93.0 & 76.1 & 44.7 & 47.3 & 36.8 & 21.5 \\
+AGR (MSE) & 93.7 & 76.5 & 49.7 & 47.7 & 36.9 & 21.6 \\
+AGR (NLL) & 95.2 & 79.0 & 50.1 & 51.9 & 39.1 & 22.1 \\

\textbf{RAVN (Ours)} & \textbf{97.0} & \textbf{80.2} & \textbf{52.4} & \textbf{53.1} & \textbf{40.3} & \textbf{23.4} \\
\bottomrule
\end{tabular}
}
\end{table}

\begin{table}[t]
\centering
\caption{\textbf{Ablation Study on Mp3d Dataset.} We compare different variants to validate the effectiveness of our proposed modules.}
\label{tab:ablation_mp3d}
\resizebox{\linewidth}{!}{
\begin{tabular}{l|ccc|ccc}
\toprule

& \multicolumn{6}{c}{\textbf{Matterport3D}} \\
\cmidrule(lr){2-7}


\textbf{Model Variant} & \multicolumn{3}{c|}{\textbf{Heard}} & \multicolumn{3}{c}{\textbf{Unheard}} \\
\cmidrule(lr){2-4} \cmidrule(lr){5-7}

& SR($\uparrow$) & SPL($\uparrow$) & SNA($\uparrow$) & SR($\uparrow$) & SPL($\uparrow$) & SNA($\uparrow$) \\
\midrule

AV-Nav~\cite{chen2020soundspaces}  & 68.8 & 52.3 & 29.6 & 33.5 & 21.9 & 10.4 \\
+AGR (MSE) & 69.4 & 50.9 & 27.0 & 35.3 & 23.8 & 12.4 \\
+AGR (NLL) & 69.7 & 54.3 & 29.4 & 35.4 & 24.1 & 13.0 \\
\textbf{RAVN (Ours)} & \textbf{70.9} & \textbf{55.6} & \textbf{33.2} & \textbf{35.6} & \textbf{25.9} & \textbf{13.6} \\
\bottomrule
\end{tabular}
}
\end{table}

\section{Conclusion}

In this work, we introduce Reliability-Aware Audio-Visual Navigation (RAVN), a framework that mitigates acoustic ambiguity via reliability-aware geometric fusion. By integrating Acoustic Geometry Reasoning (AGR) to model uncertainty and Reliability-Aware Geometric Modulation (RAGM) to dynamically gate visual features, RAVN effectively filters out unreliable auditory cues. Extensive experiments on Replica and Matterport3D demonstrate that RAVN significantly improves navigation success and efficiency, particularly in the challenging, unheard-sound generalization setting. 
Our results confirm that explicitly modeling sensory reliability is crucial for stable AVN. 

Future work will focus on extending this framework to physical robots and explicitly evaluating its robustness under extreme conditions—such as severe environmental noise and dynamic sensor degradation. Designing these highly degraded scenarios will further validate the measurable advantages of uncertainty-aware fusion in navigating unpredictable real-world acoustics.

\section*{ACKNOWLEDGMENT}

This research was financially supported by the National Natural Science Foundation of China (Grant No.: 62463029).

\bibliographystyle{IEEEtran}    

\bibliography{IEEEabrv, myref}  

\end{document}